\documentclass[aps,prl,twocolumn,superscriptaddress,floatfix]{revtex4-1}

\usepackage{graphicx}
\usepackage{amsmath,amssymb}
\usepackage{hyperref}
\usepackage{datetime}
\usepackage[utf8]{inputenc}
\usepackage{braket}

\begin{document}

\title{Quantum critical behavior at the many-body-localization transition}

\author{Matthew~Rispoli}
\author{Alexander~Lukin}
\author{Robert~Schittko}
\author{Sooshin~Kim}
\author{ M.~Eric~Tai} 
\author{Julian~L\'eonard}
\author{Markus~Greiner}

\email{greiner@physics.harvard.edu}
\affiliation{Department of Physics, Harvard University, Cambridge, Massachusetts 02138, USA}

\begin{abstract}

\end{abstract}

\maketitle

\date{\today}

\twocolumngrid

\textbf{Phase transitions are driven by collective fluctuations of a system's constituents that emerge at a critical point \cite{Tauber2017}. This mechanism has been extensively explored for classical and quantum systems in equilibrium, whose critical behavior is described by a general theory of phase transitions. Recently, however, fundamentally distinct phase transitions have been discovered for out-of-equilibrium quantum systems, which can exhibit critical behavior that defies this description and is not well understood \cite{Tauber2017}. A paradigmatic example is the many-body-localization (MBL) transition, which marks the breakdown of quantum thermalization \cite{Basko2006, Pal2010, Serbyn2013b, Huse2014, Dalessio2016,Abanin2018, Schreiber2015, Smith2015, Choi2016, Lukin2018}. Characterizing quantum critical behavior in an MBL system requires the measurement of its entanglement properties over space and time \cite{Serbyn2013b,Huse2014,Abanin2018}, which has proven experimentally challenging due to stringent requirements on quantum state preparation and system isolation. Here, we observe quantum critical behavior at the MBL transition in a disordered Bose-Hubbard system and characterize its entanglement properties via its quantum correlations. We observe strong correlations, whose emergence is accompanied by the onset of anomalous diffusive transport throughout the system, and verify their critical nature by measuring their system-size dependence. The correlations extend to high orders in the quantum critical regime and appear to form via a sparse network of many-body resonances that spans the entire system \cite{Potter2015,Khemani2017}. Our results unify the system's microscopic structure with its macroscopic quantum critical behavior, and they provide an essential step towards understanding criticality and universality in non-equilibrium systems \cite{Tauber2017,Khemani2017,Abanin2018}.}

\begin{figure}[hb!]
	\includegraphics{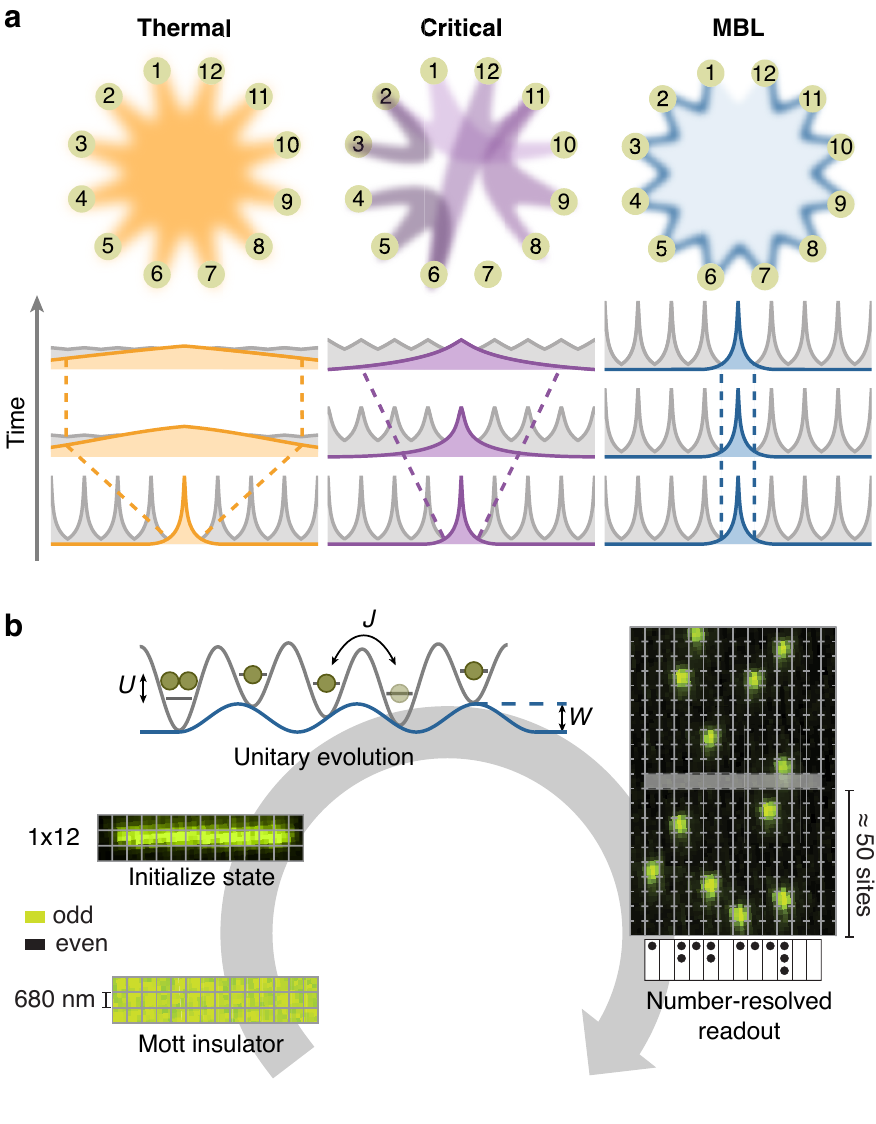}
	\caption{\textbf{Microscopy of the many-body localization transition. a,} The quantum state at the critical point takes on a complex pattern of strong multi-particle correlations at all length scales, visualized by shaded links between different lattice sites. In contrast, it simplifies in the thermal and the MBL phases to maximal entanglement and predominantly local correlations, respectively. A consequence is a change in the transport properties from diffusive to anomalous before ceasing completely in MBL. \textbf{b,} We initialize the system as a pure product state of up to twelve lattice sites at unity filling. The system becomes entangled under the unitary, non-equilibrium dynamics of the bosonic, interacting Aubry-Andr\'e model with on-site interaction energy $U$, particle tunneling at rate $J/\hbar$ (with the reduced Planck constant $\hbar$), and quasi-periodic potential with amplitude $W$. After a variable evolution time, we obtain the full atom-number distribution from site-resolved fluorescence imaging after expansion (see Methods).}
	\label{fig:fauxPhaseDiag}
\end{figure}

The many-body-localization (MBL) transition describes the breakdown of thermalization in an isolated quantum many-body system as disorder is increased beyond a critical value \cite{Schreiber2015, Smith2015, Choi2016, Lukin2018}. It represents a novel type of quantum phase transition that fundamentally differs from both its classical and quantum ground-state counterparts \cite{Basko2006, Pal2010, Abanin2018}. Instead of being characterized by an instantaneous thermodynamic signature, it is identified by the system's inherent dynamic behavior. In particular, the MBL transition manifests itself through a change in entanglement dynamics \cite{Lukin2018}. Recent years have seen tremendous progress in our understanding of both the thermal and the MBL phases within the frameworks of quantum thermalization \cite{Dalessio2016, Neill2016, Kaufman2016} and emergent integrability \cite{Serbyn2013b, Huse2014, Schreiber2015, Smith2015, Choi2016, Lukin2018}, respectively.

\begin{figure*}[ht!]
	\includegraphics{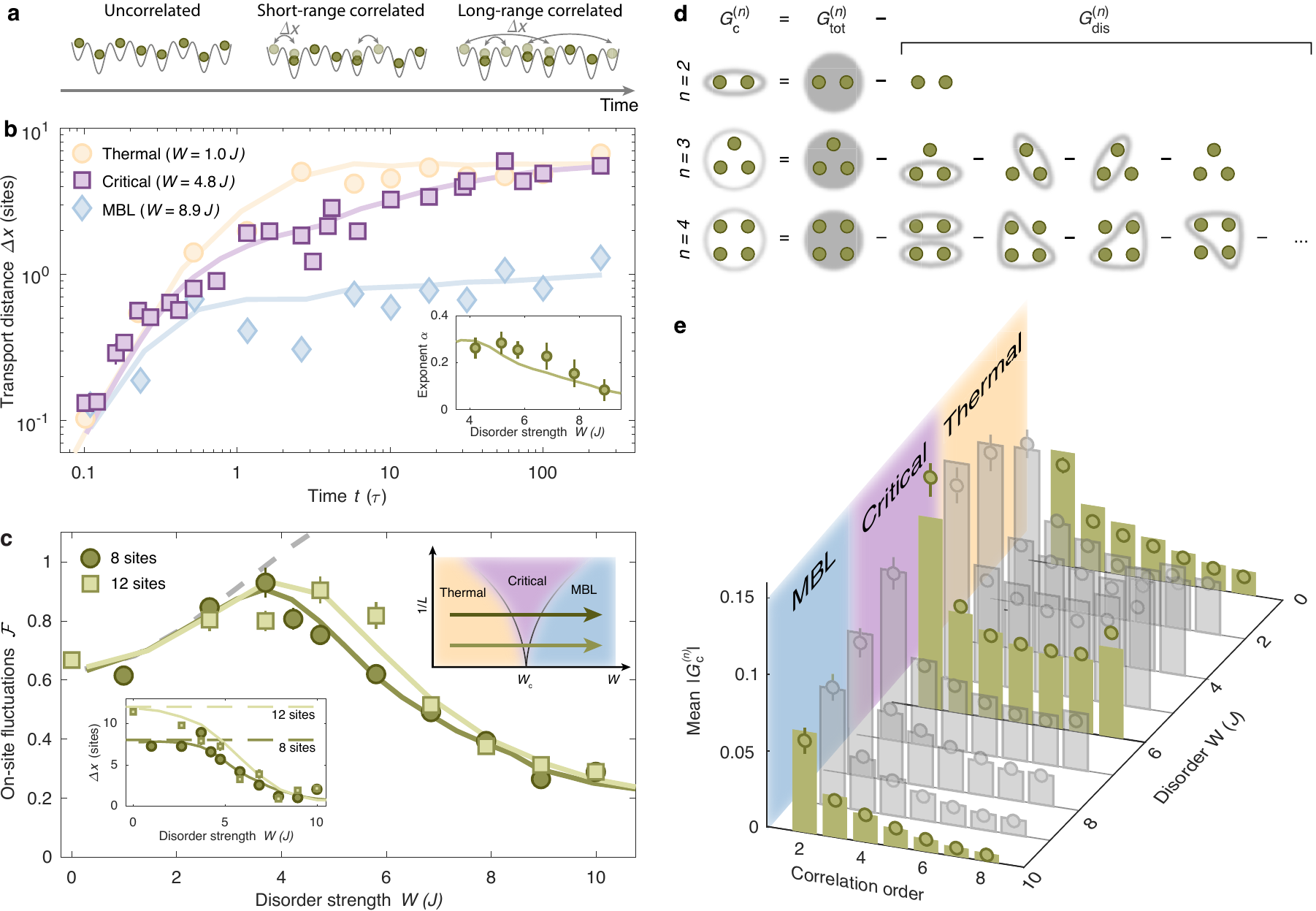}
	\caption{ \textbf{Quantum critical dynamics at the MBL transition. a,} The initially uncorrelated system develops two-point density correlations under its transport dynamics. Short-range correlations emerge within one tunneling time $\tau=\hbar/J$, whereas the diffusion exponent $\alpha$ determines the time scale over which correlations form across the system size $L$. \textbf{b,} Particle transport slows down at intermediate disorder, consistent with a power-law evolution with exponent $\alpha<0.5$, demonstrating subdiffusive dynamics (inset). These data were taken on an eight-site system. \textbf{c,} The critical nature of these dynamics is determined from the behavior of on-site density fluctuations $\mathcal{F}$ and transport distance $\Delta x$ (lower left inset) for both considered system sizes. The thermal regime is determined by the agreement of the measured $\mathcal{F}$ with the prediction from a thermal ensemble (dashed grey). The system-size dependence at intermediate disorder is consistent with the reduced size of a quantum critical cone (upper right inset). \textbf{d,} We obtain the genuine many-body processes of order $n$ from connected correlations $G_c^{(n)}$ by subtracting all lower order contributions $G_\text{dis}^{(n)}$ from the total correlation function $G_\text{tot}^{(n)}$. \textbf{e,} In the quantum critical regime, we find enhanced collective fluctuations at all measured orders by computing the mean absolute value of $G_\text{c}^{(n)}$ for different disorder strengths. The solid lines (\textbf{b,c}) and bars (\textbf{e}) denote the prediction of exact diagonalization calculations without any free parameters (see Methods). The errorbars are the s.e.m. and are below the marker size in \textbf{b}.}
	\label{fig:transport}
\end{figure*}

The quantum critical behavior at this transition, however, has remained largely unresolved \cite{Abanin2018}. In particular, it is unclear whether the traditional association of collective fluctuations with static and  dynamic critical behavior can be applied to this transition. The high amount of entanglement found at the MBL transition limits numerical studies due to the required computational power \cite{Agarwal2015, Setiawan2017}. Several theoretical approaches, despite using disparate microscopic structures, suggest anomalous transport as the macroscopic behavior at the quantum critical point \cite{Vosk2015, Potter2015, Dumitrescu2017, Goremykina2018}. Experimental studies indeed indicate a slowdown of the dynamics at intermediate disorder \cite{Lueschen2017a, Bordia2017a}. However, identifying anomalous transport as quantum critical dynamics is experimentally challenging, since similar behavior can also originate from stochastic effects: rare regions in the disorder potential \cite{Roeck2017, Nandkishore2017, Agarwal2017}, inhomogeneities in the initial state \cite{Luitz2016}, or the coupling to a classical bath \cite{Nandkishore2014, Lueschen2017}. Our experimental protocol overcomes these challenges by using a quasi-periodic potential, which is rare-region free, as well as by evolving a pure, homogeneous initial state under unitary dynamics.  Using this protocol, we observe quantum critical dynamics via anomalous transport, enhanced quantum fluctuations, and system-size dependent thermalization. In addition, we microscopically resolve and characterize the structure of the entanglement in the many-body states through their multi-particle quantum correlations. 

Our experiments start with a pure state of up to twelve unentangled lattice sites at unity filling (see Methods) and study its out-of-equilibrium evolution under the bosonic, interacting Aubry-Andr\'e Hamiltonian:

\begin{equation*}
\begin{aligned}
	\hat{\mathcal{H}} &=  -J \sum_i \left(\hat{a}_i^\dagger\hat{a}_{i+1} + h.c.\right) \\
	&+ \frac{U}{2} \sum_i\hat{n}_i \left( \hat{n}_i - 1\right) + W\sum_i h_i\hat{n}_i\text{,}
\end{aligned}
\end{equation*}
where $\hat{a}_i^\dagger$ ($\hat{a}_i$) is the creation (annihilation) operator for a boson on site $i$, and $\hat{n}_i$ is the corresponding particle number operator. The tunneling time $\tau=\hbar/J=4.3(1)\, \text{ms}$ (with the reduced Planck constant $\hbar$) between neighboring sites and the pair-wise interaction energy $U=2.87(3) J$ remain constant for all experiments. The chemical potential $h_i = \cos\left(2\pi\beta i+\phi\right)$ on site $i$ follows a quasi-periodic distribution of amplitude $W$, period $1/\beta\approx 1.618$ lattice sites, and phase $\phi$. After a variable evolution time, we obtain full counting statistics of the quantum state through a fluorescence imaging technique. The applied unitary evolution preserves the initial purity of $99.1(2)\%$ per site, such that all correlations are expected to stem from entanglement in the system \cite{Kaufman2016,Lukin2018}.

We first characterize the system's dynamical behavior by studying its transport properties for different disorder strengths. Since the initial state has exactly one atom per site, the system starts with zero density correlations at all length scales. However, during the Hamiltonian evolution, tunneling dynamics build up anti-correlated density fluctuations between coupled sites of increasing distance (Fig.~\ref{fig:transport}a). Motivated by this picture, we quantify the particle dynamics by defining the transport distance, $\Delta x = 2 \sum_{d} d \times \langle G_\text{c}^{(2)}(i,i+d) \rangle_i$, as the first moment of the disorder-averaged two-point density correlations, $G_\text{c}^{(2)}(i,i+d)=\langle \hat{n}_i \hat{n}_{i+d} \rangle - \langle \hat{n}_i \rangle \langle \hat{n}_{i+d} \rangle$ (Fig.~\ref{fig:transport}a). With increasing disorder, we observe a slowdown of particle transport that is consistent with a power-law growth $\Delta x \sim t^\alpha$ (Fig.~\ref{fig:transport}b, see Methods) \cite{Lucioni2011}. We extract the anomalous diffusion exponent $\alpha$ from a subset of the data points which excludes the initial transient dynamics (Fig.~\ref{fig:transport}b inset). The exponent $\alpha$ approaches zero for successively higher disorder, demonstrating the absence of transport in the MBL regime.

\begin{figure}[ht!]
	\includegraphics{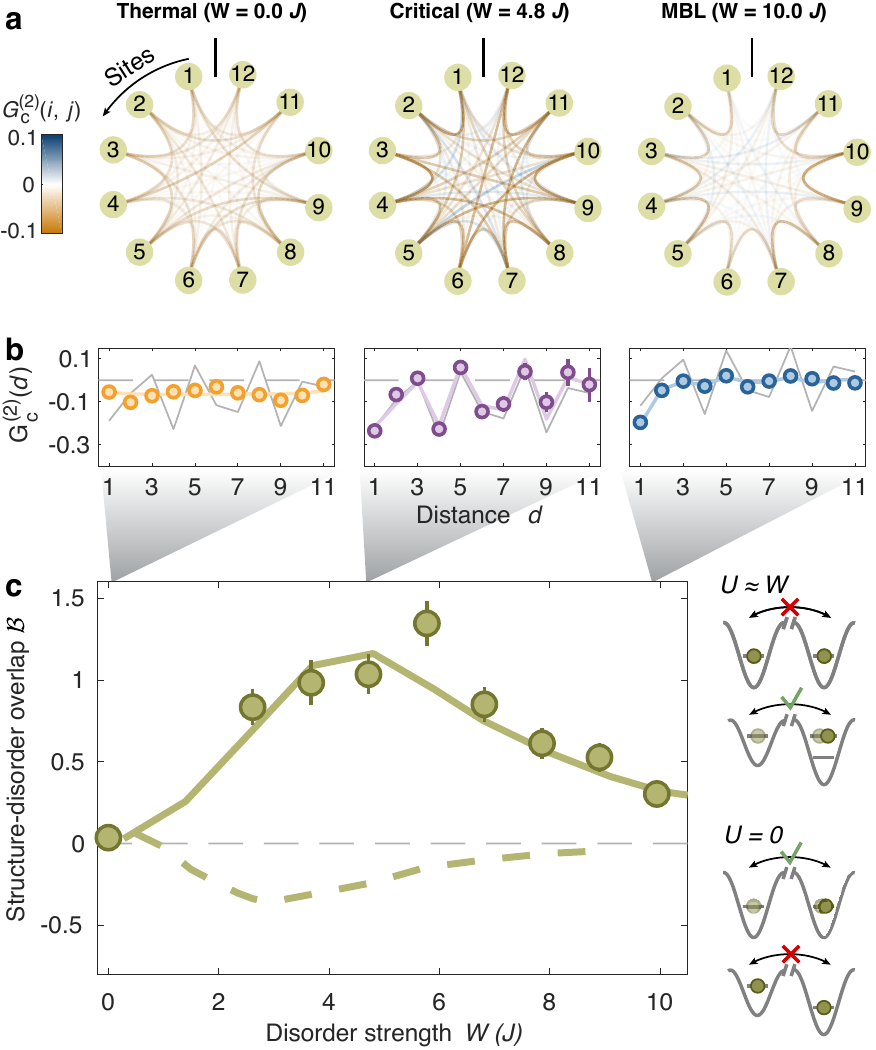}
	\caption{\textbf{Sparse network of resonances. a,} The measured site-dependent two-point correlations $G_\text{c}^{(2)}(i,j)$ qualitatively differ for the three disorder regimes. In the quantum critical regime, correlations preferably form at specific distances, showing a network-like structure. This contrasts with homogeneous correlations in the thermal regime and nearest-neighbor correlations in the MBL regime. \textbf{b,} The structure of the correlation network is revealed by the averaged correlation function $G_\text{c}^{(2)}(d) = \langle G_\text{c}^{(2)}(i,i+d)\rangle_i$. Its similarity to the autocorrelation $A(d) = \langle h_i h_{i+d}\rangle_{i}$ of the quasi-periodic potential (solid grey) indicates interaction-induced tunneling processes that are enhanced when the interaction energy compensates for the chemical potential difference. \textbf{c,} We quantify the similarity by the overlap $\mathcal{B} = \Sigma_d G_\text{c}^{(2)}(d)  A(d)$, which is maximal in the quantum critical regime. The sign of the overlap would be opposite for non-interacting particles (dashed line), which favors tunneling between sites with similar potential energies. The solid lines in \textbf{b,c} denote the prediction of exact diagonalization calculations without any free parameters. The errorbars are the s.e.m.}
	\label{fig:g2}
\end{figure}

In order to identify the anomalous diffusion as a signature of quantum critical dynamics, we measure the system-size dependence of two observables in the long-time limit $(t=100\tau)$: the on-site number fluctuations $\mathcal{F} \equiv G_\text{c}^{(2)}(d=0)$ as a probe of local thermalization, and the transport distance $\Delta x$ as a localization measure (Fig.~\ref{fig:transport}c, see Methods). At low disorder, the fluctuations agree with those predicted by a thermal ensemble and particles are completely delocalized for both system sizes. This demonstrates that local quantum thermalization in our system is not subject to finite-size effects. At strong disorder, we find sub-thermal fluctuations and a transport distance $\Delta x \ll L$. This indicates that the physics is governed by a system-size independent, intrinsic length scale, namely the localization length $\Delta x$ \cite{Choi2016, Lukin2018}. However, at intermediate disorder, we find a system-size dependence for both observables, demonstrating the absence of an intrinsic length scale and the presence of finite-size-limited local fluctuations. Our measurements can be visualized as two horizontal cuts in a finite-size phase diagram, whose finite-size dependence agrees with the physics associated with a shrinking quantum critical cone  (Fig.~\ref{fig:transport}c inset) \cite{Tauber2017}.

We then investigate the multi-particle correlations in the system to probe the presence of enhanced quantum fluctuations in the quantum critical regime (Fig. \ref{fig:transport}d). For this study, we employ the $n$-point connected density-correlation functions \cite{Liu2016, Schweigler2017,Hodgman2017},
\begin{equation*}
G_\text{c}^{(n)}(\textbf{x}) = G_\text{tot}^{(n)}(\textbf{x}) - G_\text{dis}^{(n)}(\textbf{x}),
\end{equation*}
which act on lattice sites with positions $\textbf{x} = (x_1,...,x_n)$. The disconnected part of this function, $G_\text{dis}^{(n)}$, is fully determined by all lower-order correlation functions, and therefore does not contain new information at order $n$. By removing it from the total measured correlation function, $G_\text{tot}^{(n)}$, we isolate all $n$-order correlations that are independent of lower-order processes (see Methods). This approach gives a direct handle on the level of complexity of the underlying many-body wave function and characterizes its non-seperabilty into sub-systems of size $<n$. We quantify the relevance of order $n$ processes by computing the mean absolute value of all correlations arising from both contiguous and non-contiguous $n$ sites in the system (Fig.~\ref{fig:transport}e). We find that in the thermal and the many-body-localized regimes, the system becomes successively less correlated at higher order. The behavior in the quantum critical regime is strikingly different: we observe that the system is strongly correlated at all measured orders.

In order to reveal the microscopic origin for the anomalous transport, we now investigate the site-resolved structure of the many-body state (Fig.~\ref{fig:g2}a). We first study how much each lattice site contributes to the transport by considering the site-resolved two-point correlations in the long-time limit. In the thermal regime, we find similar correlations between all lattice sites, which correspond to uniformly delocalized atoms. In contrast, density correlations are restricted to nearby sites in the MBL regime due to localization. Intriguingly, we observe a sparse structure of correlations at intermediate disorder, which involves only specific distances between lattice sites, yet spans the entire system size. 

The sparse structure is expected to be linked to the applied quasi-periodic potential. The average energy offsets of sites $d$ apart in the system are correlated by this potential. This correlation is then inherited by the system's fluctuations when the interaction energy $U$ compensates for these correlated offsets. To investigate this structure, we compare the two-point density correlations with the autocorrelation function, $A(d)=\langle h_i h_{i+d}\rangle_i$, of the quasi-periodic potential. Indeed, we find that the site-averaged density correlations $G_\text{c}^{(2)}(d) = \langle G^{(2)}(i,i+d)\rangle_i$ inherit their spatial structure from $A(d)$ (Fig.~\ref{fig:g2}b). We find that this contribution is maximal in the critical regime but is strongly reduced in the thermal and MBL regimes (Fig.~\ref{fig:g2}c). These observations contrast with the behavior of a non-interacting system, where the sign of the structure is opposite since resonant tunneling is favored for zero potential energy difference (Fig.~\ref{fig:g2}c). These results illustrate microscopically how the interplay of strong interactions and disorder can lead to anomalous diffusion. However, this picture of effective single-particle hopping that couples distant sites neglects the many-body nature of these systems.

\begin{figure}[t]
		\includegraphics{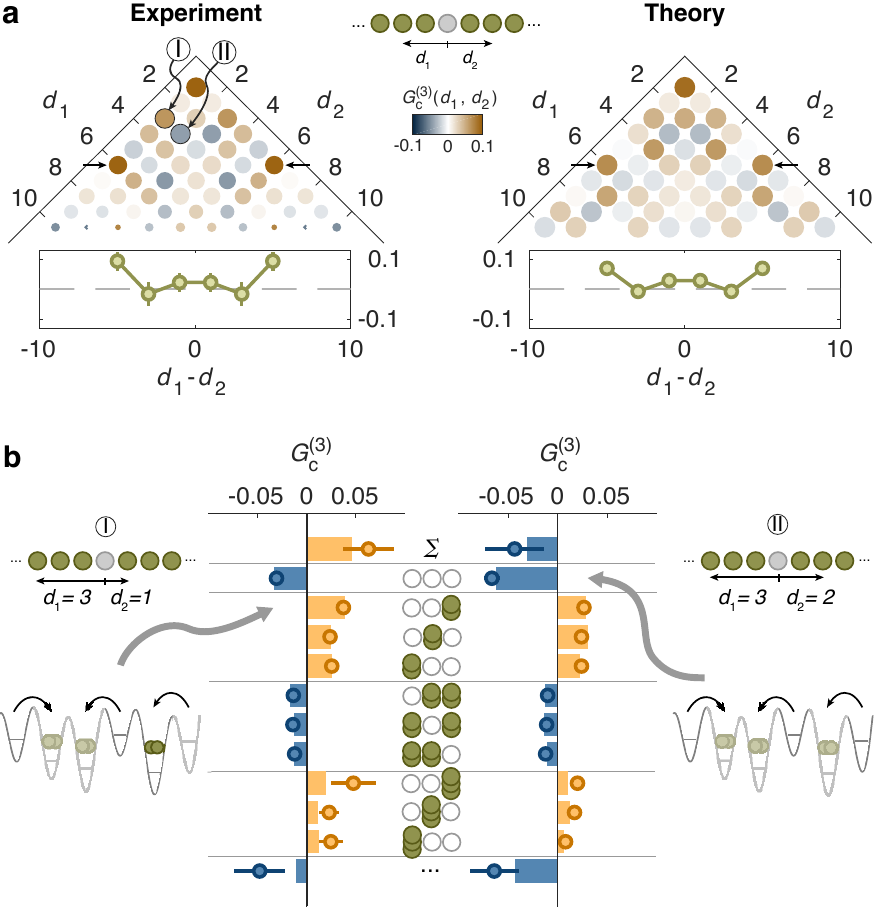} 
	\caption{\textbf{Many-body correlations in the quantum critical regime. a,} The connected correlation function, $G^{(3)}_c(d_1,d_2)$, for three lattice sites spaced by distances $d_1$ and $d_2$ in the quantum critical regime $(W=4.8J)$, showing the strongly interacting nature of the state. We find that the three-point correlations show a characteristic structure that is governed by the contribution of the number states on the considered sites. The arrows indicate the cut in $d_1,d_2$ space plotted below. \textbf{b,} To exemplify the relevant processes of order n=3, we show the contributions of the number states on lattice sites at distance $d_1=3$, $d_2=1$ (left) and $d_1=3$, $d_2=2$ (right). While there is a wide distribution of contributing configurations, the relative dominance of a particular process provides the overall structure in \textbf{a}. The cartoons illustrate how a highly correlated hopping process can give rise to positive or negative correlations depending on the three considered sites. The theory plot in \textbf{a} and bars in \textbf{b} are calculated from exact diagonalization without any free parameters. The inverse marker size in \textbf{a} and the error bars denote the s.e.m.}
		\label{fig:gn}	
\end{figure}

In order to investigate the system's many-body structure, we examine the site-resolved contributions of the three-point correlations. Since all non-zero contributions to the three-point correlations involve correlated hopping of at least two particles, they are a signature for multi-particle entanglement and therefore demonstrate a breakdown of mean-field approximations \cite{Schweigler2017}. In the quantum critical regime, we find that these correlations span the entire system and are highly structured, taking on both positive and negative values (Fig.~\ref{fig:gn}a). In contrast to the pattern in the second-order correlation function, this third-order structure is not directly recognizable as the quasi-periodic-potential correlations. In order to gain further insight into the structure, we analyze the contributions of all possible particle configurations in Fig.~\ref{fig:gn}b. In particular, for $G^{(3)}_c(d_1=3,d_2=1)$, which is positive, we see that the dominant contribution comes from a particular process that favors multiple atoms hopping to the same site. In contrast, $G^{(3)}_c(d_1=3,d_2=2)$, which is negative, has a dominant process that favors all atoms leaving the three sites considered. While this provides some intuition for the emergent many-body resonances, the three-point correlations are, in fact, the result of a superposition of many correlated processes. These observations further demonstrate how the interactions between multiple atoms can compensate for the disorder via correlated tunneling of several atoms. In this way, we can see the additional role interactions play in the disordered system: they supply higher-order many-body resonances that preserve transport where lower-order processes are energetically suppressed.

Our results demonstrate how a many-body, sparse resonant structure drives the quantum critical behavior at the MBL transition. This observed microscopic description is consistent with the theoretically suggested mechanisms of a sparse \textit{backbone} of resonances that can act as a functional bath for the system \cite{Potter2015, Khemani2017}. However, our results provide a new perspective on this description by mapping out the prevalence of high-order processes in the system that facilitate this critical thermalization.

In future experiments, the tunability of our system will allow us to address further open questions on the MBL transition, such as possible discontinuities of the entanglement entropy \cite{Khemani2017}, the potential emergence of new dynamic phases near the critical point, and the influence of rare-regions in the disorder potential \cite{Roeck2017, Nandkishore2017}. Furthermore, the demonstrated techniques pave the way to explore the role of universality in non-equilibrium systems. From a quantum computing perspective, our system's Hilbert space dimension exceeds the dimension of 22 spins with zero total magnetization, bringing numerically intractable sizes within experimental access. 

\begin{acknowledgements}
We acknowledge fruitful discussions with D.~Abanin, E.~Altman, H.~Bernien, C.~Chiu, S.~Choi, E.~Demler, A.~H\'ebert, M.~Heyl, W.~W.~Ho, V. Kasper, V.~Khemani, J.~Kwan, D.~Luitz and L.~Santos. We are supported by grants from the National Science Foundation, the Gordon and Betty Moore Foundations EPiQS Initiative, an Air Force Office of Scientific Research MURI program, an Army Research Office MURI program and the NSF Graduate Research Fellowship Program. J. L. acknowledges support from the Swiss National Science Foundation. 
\end{acknowledgements}

\bibliography{mbl}


\onecolumngrid

\section{Supplementary information}
\subsection{Experimental protocol}

All of our experiments start with a unity-filling, two-dimensional Mott insulator of bosonic $^{87}$Rb atoms in a deep, blue-detuned optical lattice with lattice constant $a=680\text{nm}$ and $45E_\text{r}$ lattice depth, where $E_\text{r}=h \times 1.1\,\text{kHz}$ is the recoil energy for an atom with mass $m$, and $h$ is the Planck constant. We then employ two digital micromirror devices (DMDs) in the Fourier plane of our microscope to optically confine a single chain of $N$ ($N = 8,12$) atoms chosen from the Mott insulator's unity-filling shell and subsequently ramp down the power of the optical lattice. Upon waiting for a few tens of milliseconds to allow for a departure of the non-confined atoms surrounding the chain, we then turn on the lattice again and ramp down the confining DMD potentials, thereby finalizing the initialization of our unity-filling, 1x$N$ system. Taking into account the additional post-selection described below, this procedure results in a 99.1(2)\% probability of finding exactly one atom on any given site. Notably, this value establishes lower bounds of 93\% and 90\% for the global quantum state purities of the 8-site and the 12-site system, respectively. 

Upon initializing the system of our choice, we subsequently pursue three separate paths of action to initiate the dynamics we are interested in studying. In a first step, we use one of our DMDs to project an optical potential onto our atoms. This ``wall-potential'' provides a box-like confinement which is registered to the position of the optical lattice, and later defines the size of the one-dimensional system once the bare lattice depth has been lowered. Secondly, we simultaneously use the other DMD to project a custom, quasi-periodic disorder potential onto our atoms. Finally, after both of these potentials have been turned on, we rapidly lower the bare lattice depth along the atomic chain from  $V_x = 45E_\text{r}$ to $V_x = 8E_\text{r}$, thereby quenching the system and giving rise to many-body dynamics. After a variable evolution time in this lowered potential, we freeze said dynamics by quickly ramping the longitudinal lattice back up to $V_x = 45E_\text{r}$. 

We then let the atom populations located on individual lattice sites expand into independent tubes and use fluorescence imaging to perform a site-resolved atom number measurement. The expansion step before the imaging procedure is employed to avoid parity projection during the imaging process. We subsequently post-select our data by excluding any images which do not contain the correct total number of atoms. \\

The steps briefly described above are conceptually identical to those employed in \cite{Kaufman2016,Lukin2018}, where they are described in more detail in the methods sections.

\subsection{High-Order Correlation Functions}

Generically, a $n^\mathrm{th}$ order correlation function can be measured from a set of operators $\mathcal{O}_i$ by their joint expectation value $\langle \prod_{i=1}^n \mathcal{O}_i \rangle = \langle \mathcal{O}_1 \mathcal{O}_2 ... \mathcal{O}_n \rangle$. However, this joint expectation value captures two kinds of information: ``disconnected'' correlations that exist at $n^\mathrm{th}$ order due to existing lower order correlations, and ``connected'' correlations that only exist at order $n$ and can't be described by factorization into correlations of lower order \cite{Liu2016,Schweigler2017,Hodgman2017,Preiss2018}.
 
In the two-point case, this would mean comparing the measured value of $\langle \mathcal{O}_i \mathcal{O}_j \rangle$ to the product of their individual expectation values $\langle \mathcal{O}_i \rangle \langle \mathcal{O}_j \rangle$. The ``connected'' part of the correlation between $i$ and $j$ is  defined as the correlations that remain after removing the contributions from factorization into smaller groups. This motivates the definition of $G_\text{c}^{(2)}(i,j)=\langle \mathcal{O}_i \mathcal{O}_j \rangle - \langle \mathcal{O}_i \rangle \langle \mathcal{O}_j \rangle$.
 
 To provide some intuition, we describe two concrete examples in terms of two-point joint expectation values constructed from atom-number operators $\hat{n}_i$ as $\langle n_1 n_2 \rangle$. The two example states are:
 \begin{flalign*}
| \psi_1 \rangle = &{}\frac{1}{\sqrt{2}} \left ( |0 \rangle + |2 \rangle \right ) \otimes \frac{1}{\sqrt{2}} \left ( |0 \rangle + |2 \rangle \right ) \\
| \psi_2 \rangle =&{} \frac{1}{\sqrt{2}} \left ( |2 2 \rangle + | 0 0 \rangle \right ) 
\end{flalign*}

 For both $ | \psi_1 \rangle$ and $ | \psi_2 \rangle$   we see that they have the same local fluctuations $\langle n_i^2 \rangle - \langle n_i \rangle^2 = 1$ and local on-site number $\langle n_i \rangle = 1$. However, we see, by construction, that $| \psi_1 \rangle$ is created from an outter product of states with these local fluctuations and therefore should have uncorrelated joint fluctuations.   This means that sites would give the joint expectation value of two incoherently fluctuating random variables, i.e. $G_\text{c}^{(2)}(|\psi_1\rangle) = 1 - 1 = 0$ or site-1 and site-2 in $| \psi_1\rangle$ have no genuine ``connected'' two-point correlations. This differs from the second case of $| \psi_2 \rangle$ where $ \langle n_{i} \rangle=1$  and $\langle n_i^2 \rangle - \langle n_i \rangle^2 = 1$  but the correlated fluctuations of the value bring $\langle n_1 n_2 \rangle = 2$ which then gives a ``connected'' correlation value of $G_\text{c}^{(2)}(|\psi_2\rangle)=2 - 1 =1$. This shows that the state $|\psi_2\rangle$ has genuine two-point correlations that cannot be described by factorization into smaller groups.

For a three-point ``connected'' correlation function, we must subtract out contributions that come from ``connected'' two-point correlations that can look like three-point correlations when randomly combined with a residual 1-point correlation. This is how the ``connected'' three-point correlation function is defined in the main text as $G_\text{c}^{(3)}$ for the on-site number operator $\hat{n}_i$. 
 \begin{flalign*}
G_\text{c}^{(3)}(i,j,k) = &{} \langle \mathcal{O}_i \mathcal{O}_j \mathcal{O}_k \rangle \\
 - &{} G_\text{c}^{(2)}(i,j) \langle \mathcal{O}_k \rangle - G_\text{c}^{(2)}(i,k) \langle \mathcal{O}_j \rangle  - G_\text{c}^{(2)}(j,k) \langle \mathcal{O}_i \rangle \\
      - &{} \langle \mathcal{O}_j \rangle \langle \mathcal{O}_j \rangle \langle \mathcal{O}_k \rangle
\end{flalign*}

This leads to a recursive definition of $n$-order ``connected'' correlation functions that then depend on the integer partitioning and permutations of all lower-order ``connected'' correlations. For convenience and compact notation, we will make a set of useful definitions:
\begin{itemize}
\item $G_\text{c}^{(1)}(i) = \langle \mathcal{O}_i \rangle$ \& $G_\text{c}^{(0)}(\emptyset)=1$
\item a function $\hat{\mathcal{P}}_{{\{ i_1,i_2,...,i_n\}}\choose{\{k\in g\}}} \big ( \hspace{0.5mm} ... \hspace{0.5mm}  \big )$ that finds all unique permutations of choosing all indices in the group $\{ i_1, i_2, ..., i_n \}$ separated into integer partitions $\{k\}=\{k_1,k_2,...,k_g\}$ that sum to $n$: $\sum_i^g k_i = n$
\item  $\sum_{g \in N}$ that defines a sum over all unique integer partitions $\{g\}$ of the order $n$ 
\end{itemize}

 \begin{equation*}
G_\text{c}^{(n)}(i_1,i_2,...,i_n) = \langle \mathcal{O}_{i_1} \mathcal{O}_{i_2} ... \mathcal{O}_{i_n} \rangle  - \sum_{g \in N} \left [ \hat{\mathcal{P}}_{{\{ i_1,i_2,...,i_n\}}\choose{\{k\in g\}}} \left ( \prod_{k_i}^g G_\text{c}^{(k_i)} \left [ {{\{ i_1,i_2,...,i_n\}}\choose{\{k_i\in k\}}}\right ] \right ) \right ]
\end{equation*}

\vspace{10px}
\begin{figure}[t]
	\includegraphics[width=\columnwidth]{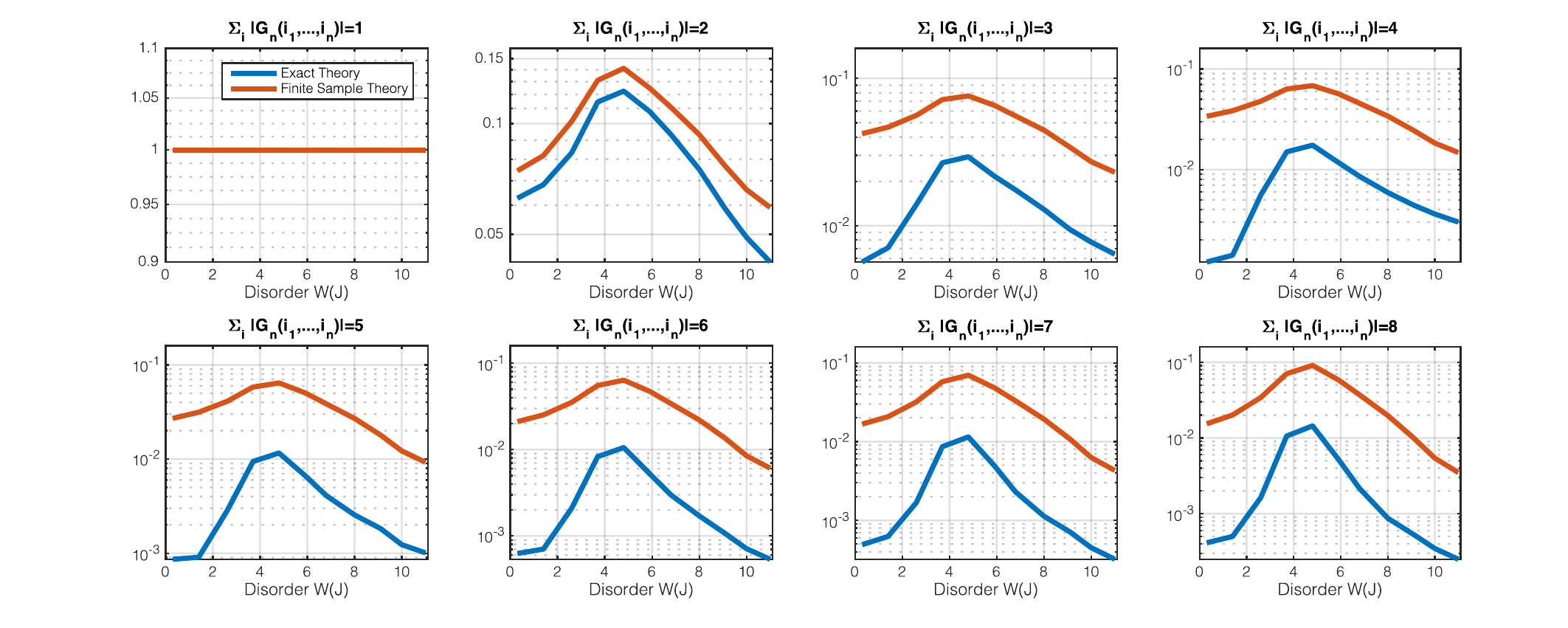}
	\caption{\textbf{Finite Sampling Bias for high-order correlations} Each of these panels compare the theory for the average absolute value of the high-order correlations measured at  $t=100 \tau$. The exact theory (blue)  was sampled at the same rate as the experimental data taken to produce the Monte-Carlo-sampled theory (orange) which exhibits a systematic upwards bias in this absolute-value measure. The qualitative trend of the critical regime having the highest correlations, however, remains unaffected.
	}
	\label{fig:finiteMC}
\end{figure}

In general, we must subtract all permutations of factorable groups at lower-order. This is the same as combining the integer partition problem and then finding all permutations of choosing those integer partitions from the number of indices equal to the sum. To exemplify how this looks for the next order,  it is applied to finding all the correct factorizations of the four-point correlation function. The unique integer partitions of 4 with at least two non-zero integers are $\left \{ (3,1),(2,2),(2,1,1),(1,1,1,1) \right \}$.
\begin{flalign*}
G_4(i,j,k,l) =\langle \mathcal{O}_i \mathcal{O}_j \mathcal{O}_k \mathcal{O}_l \rangle  \\
 - &{} \Big \{ G_3(j,k,l) G_1(i)- G_3(i,k,l)  G_1(j) \rangle - G_3(i,j,l) G_1(k) \rangle - G_3(i,j,k)  G_1(l) \Big \}\\
 - &{} \Big \{ G_2(i,j) G_2(k,l) - G_2(i,k) G_2(j,l) - G_2(i,l) G_2(j,k) \Big \}\\
 - &{}\Big \{  G_2(i,j) G_1(k) G_1(l) -  G_2(i,k) G_1(j) G_1(l) -  G_2(i,l) G_1(j) G_1(k) \\
  - &{}  G_2(j,k) G_1(i) G_1(l)  -  G_2(j,l) G_1(i) G_1(k)  -  G_2(k,l) G_1(i) G_1(j) \Big \}  \\
      - &{} \Big \{ G_1(i) G_1(j) G_1(k) G_1(l) \Big \}
\end{flalign*}

\subsection{Distribution of n-point correlation functions}

The probability distribution of the n-point correlation for each set of lattice sites is intrinsically asymmetric towards larger absolute values. While this bias is incorporated in the s.e.m. for a single correlation value, this results in a finite-sampling bias when calculating the mean-absolute-value. Calculating this quantity for a finite sample number therefore overestimates the expectation value in the limit of infinite sample number. We can include this effect in theory by Monte-Carlo sampling from the exact diagonalization calculations with the same number of measured shots ($\sim 110 / \mathrm{disorder}$).  This finite-sample theory is plotted in Fig.~2e. In order to ensure that the qualitative feature of enhanced correlations remains unchanged for larger sample numbers, we compare the finite-sample theory with the exact calculations, see Fig.~\ref{fig:finiteMC}. We find that indeed the two theory curves are in qualitative agreement. 

\begin{figure}[t]
	\includegraphics[width=\columnwidth]{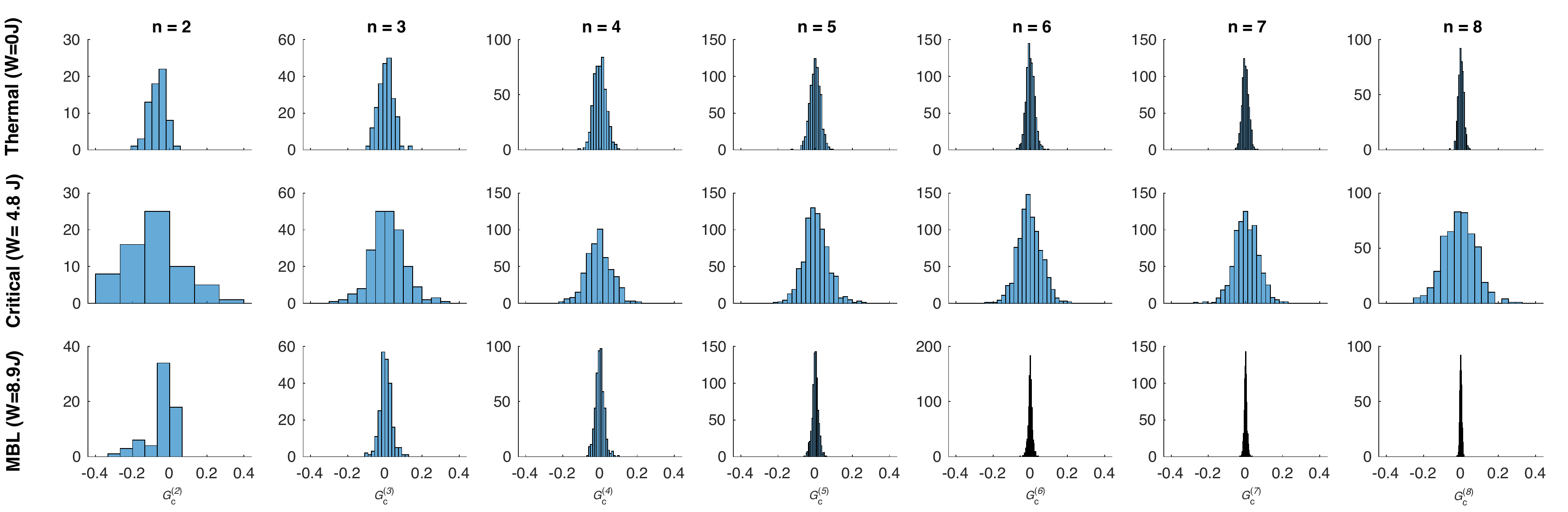}
	\caption{\textbf{Probability distribution of the n-point correlations} The histograms show the distributions from which the data in Fig.~2e (main text) are calculated. In the critical regime, we find a broad distribution of the correlations at high orders, whereas in both the thermal and the MBL regime the correlations are close to zero for all sets of lattice sites.
	}
	\label{fig:histograms}
\end{figure}

We further investigate the connected n-point correlations functions by calculating histograms of the distribution of all sets of lattice sites, see Fig.~\ref{fig:histograms}. 

\subsection{Thermal Ensemble Calculation}
At $t=0$, the system is quenched to the Hamiltonian $H_q(W)$, which depends on disorder strength $W$. The thermal prediction shown in Fig.~2c is calculated using a canonical ensemble, in which the eigenstates of the system are exponentially populated according to $P_i \sim e^{E_i/T}$, where $E_i$ is the eigenenergy of state $i$. The effective temperature of the system is determined by finding the canonical temperature $T(J)$ that yields the correct average energy $\langle E \rangle$ after quenching into the Hamiltonian $H_q(W)$. We have additionally compared this canonical ensemble to a microcanonical one, finding excellent agreement between the two. The microcanonical ensemble we used is composed of an equal-probability statistical mixture of those 11 eigenstates of $H_q(W)$ which are closest to the average energy of the initial state, which is given by $E_0=\bra{\psi_0}H_q(W)\ket{\psi_0}$. We additionally verified that these results do not depend on the exact number of included eigenstates.

\subsection{Numerics}
In order to get theoretical predictions for 8-site systems, we perform numerical exact-diagonalization (ED) calculations. For 12-site systems, matrix diagonalization is computationally challenging due to the large Hilbert space dimension. Instead, we implement an exact numerical integration of Schr\"{o}dinger's equation $\ket{\psi(t)}=e^{-iHt/\hbar}\ket{\psi_0}$ based on the Krylov-subspace method. Since the Hamiltonian is sparse, the Krylov-subspace method provides an efficient (both in terms of memory and CPU-run-time usage) way to numerically compute the time evolution while achieving high, controlled precision. All data points are averaged over 200 different disorder realizations. The computations are performed on the Harvard Odyssey computing cluster. For specifications see: https://www.rc.fas.harvard.edu/odyssey/

\vspace{10px}

\subsection{Determination of Time Scaling Behavior (Power Law 12-Sites)}

\begin{figure}[t]
	\includegraphics[width=\columnwidth]{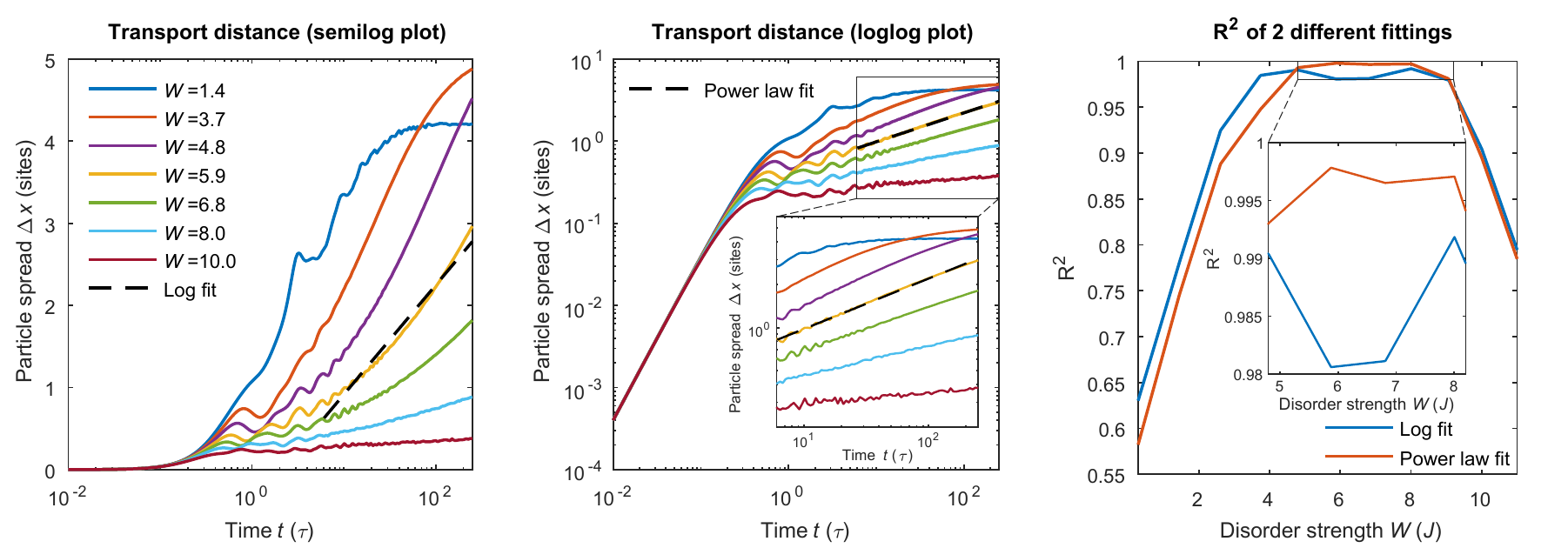}
	\caption{\textbf{Two different fittings for the time-dependence of the transport distance at various disorder strengths} The left panel and the center panels show the time dependence of the particle spread $\Delta x$ in a 12-site system for various disorder strengths in semi-log and log-log scales. The dashed lines are sample least-square fits assuming logarithm and power-law time scaling, respectively. The right panel shows the R$^2$ values of each fit as a function of disorder strength. All simulations were performed by exact numerical integration of Schr\"{o}dinger's equation, as described in the previous section.
	}
	\label{fig:TimeScalingBehavior}
\end{figure}

In order to quantify the transport distance of the particles, we defined the first moment of the two-point-density-correlation distribution, $\Delta x = \sum_d G_2(d) \times  d$ where $G_2(d) = \langle n_i n_{i+d} \rangle_{i,\phi} - \langle n_i \rangle_{i,\phi} \langle n_{i+d} \rangle_{i,\phi}$, $i$ is the site index, and $\phi$ is the disorder realization. Numerical simulations were performed to determine time-scaling behavior of the first moment (Fig.~\ref{fig:TimeScalingBehavior}), and it appears that in the critical disorder regime ($5<W/J<8$), the transport distance follows a power-law time scaling rather than a logarithmic growth. Hence, in this paper, we assumed a power-law time scaling of the particle spread to analyze the data.

\vspace{10px}

\subsection{Data Analysis}
For all experimental data we use 197 unique disorder patterns, each defined by a different phase $\phi$ of the quasi-periodic potential, and perform a running average over them by randomly sampling a given number of realizations and treating them as independent measurements of the same system. 

We extract the anomalous diffusion exponent $\alpha$ in Fig.~2b in a fit-free manner as follows: we first exclude the data at times $t<L/2\tau$, where the initial transient dynamics are still ongoing. We then calculate the slopes between all successive pairs of data points. The exponent $\alpha$ then corresponds to the average of those slopes.

The single-site atom number fluctuations $\mathcal{F}=G_\text{c}^{(2)}(d=0)$ are extracted from the edge sites for both system sizes. The edge sites are most insensitive to the introduction of additional (bulk) sites into the system and therefore allow for the fairest comparison between different total system sizes \cite{Khemani2017}. 

The number of samples for each experiment is summarized in the following table:

\begin{center}
	\begin{tabular}{|c|c|c|}
		\hline Figure & Sample number/point\\
		\hline 2B & W=1J : 153(13) / W=4.8J : 170(28) / W=8.9J : 138(3)\\ 
		\hline 2C & L=8 : 160(10) /  L=12 : 123(3) \\ 
		\hline 2E & 123(3) \\ 
		\hline 3A/B & W=0J : 424 / W=4.8J : 142 / W=9.9J : 126\\ 
		\hline 3C & 123(3) \\ 
		\hline 4 & W=4.8J : 142 \\
		\hline
	\end{tabular} 
\end{center}

\end{document}